\documentclass[11pt]{article}
\usepackage{graphicx}
\begin{document}
\title{Evidence for layered symmetry breaking in SU(2) lattice gauge theory}
\author{Michael Grady\\
Department of Physics\\ State University of New York at Fredonia\\ 
Fredonia, NY 14063 USA}
\date{\today}
\maketitle
\thispagestyle{empty}
\begin{abstract}
Simulations of four-dimensional SU(2) lattice gauge theory 
are performed with partial
axial gauge fixing trees spanning three of the four dimensions. The remaining 
SU(2) gauge symmetry, global in three directions and local in one, is found to
break spontaneously at weak coupling, with the average fourth-dimension-pointing
link in each perpendicular hyperplane as order parameter. 
The symmetry is restored at strong coupling. Symmetry breaking 
in each hyperplane appears to be independent, and occurs 
regardless of boundary conditions.
The associated phase transition is likely coincident 
with the Polyakov loop transition.  

\end{abstract}

\section{Introduction}
This paper presents evidence for a hidden symmetry breaking 
in 4-d SU(2) lattice gauge theory, 
made visible by a partial axial gauge fixing. The symmetry 
breaking occurs for weak coupling
and the symmetry is restored at strong coupling, the two 
phases being separated by a phase transition.
This transition is studied both on the Wilson axis and in the 
fundamental-adjoint plane\cite{FA} at
an adjoint coupling of $\beta _A = 1.5$. In the latter case, 
the transition is coincident with
the previously-known first-order transition seen in the 
average plaquette, and also appears
coincident with the Polyakov-loop deconfinement transition.
The observed symmetry breaking
provides an order parameter for
this transition which remains valid with open boundary conditions,
for which the Polyakov 
loop symmetry is absent.  It also shows this bulk transition to
be of the symmetry-breaking type. 
Symmetry-breaking bulk first-order transitions end in 
tricritical points rather than critical
points, because a symmetry breaking transition must 
continue into the phase diagram in order to
divide it completely into broken and unbroken regions which are not 
analytically connected.  Indeed, evidence is found of a 
higher-order phase transition, with the
same order parameter, on the Wilson axis in the crossover 
region. The existence of the bulk tricritical 
point to which this transition must attach, and the bulk 
nature of the order parameter (independent
of boundary conditions) is strong evidence for this also being a 
bulk transition. On the Wilson axis, this is
either coincident with or lies close to the Polyakov-loop 
transition on moderate size lattices up
to $16^4$. Preliminary finite size scaling evidence favors the 
transition on the Wilson axis
being of infinite-order, i.e. characterized by an essential 
singularity similar to the 
Berezinskii-Kosterlitz-Thouless (BKT) transition in the 
two-dimensional X-Y model\cite{BKT}. 
It differs from this 
transition in having a symmetry-breaking order-parameter, 
made possible by the system
having more than two dimensions.
Below such a transition, the system is critical at every 
coupling, which is the expected behavior
if the weak 
coupling phase is a massless-gluon phase characterized by 
an infinite-range force.

The existence of a bulk phase transition has important 
consequences for both 
the lattice and continuum
theories.  Simulations are relevant to the continuum theory 
only if they are run on the weak side
of such a transition. Another alternative would be to identify 
the lattice artifacts which are
responsible for the transition and devise a modified action that 
prohibits these artifacts or
at least suppresses them enough to prevent the transition 
from occurring.  
Preliminary trials indicate that the system remains in the 
broken phase for all couplings with
an action which prohibits both the SO(3)-Z2 monopoles 
defined in \cite{SO3Z2} as well as Z2
vortices (positive plaquette restriction).
If the continuum
limit is not confining, as confirmation of a bulk phase 
transition would imply, then
it would make sense to suspect the quark sector of QCD as 
the true cause of confinement, perhaps
as a side-effect of chiral symmetry breaking\cite{quark}.

\section{Motivation}
The original motivation for this study was an apparent
paradox in the 4-d Z2 lattice gauge theory. This theory
is connected to SU(2) in the fundamental-adjoint plane \cite{FA} 
in the limit of infinite adjoint coupling.
The Z2 theory undergoes a bulk first-order transition which is 
deconfining. The weak-coupling limit of 
this abelian theory is definitely not confined (no abelian 
theories are). The critical point is known
from self-duality to be exactly 
$\beta_c = \frac{1}{2} \ln (1+\sqrt{2}) \simeq 0.44$ \cite{Wegner}.
The Polyakov loop serves as an order parameter for
confinement, therefore it appears to be a proper order 
parameter for this transition. However, it,
and the symmetry broken by it, are only defined with periodic 
boundary conditions. A bulk transition, on
the other hand, should be independent of which boundary conditions 
are chosen. A symmetry-breaking 
first-order transition has a much different Landau free 
energy (even orders to sixth order present, with
negative fourth-order term) than its non-symmetry-breaking 
cousin (third-order term driving transition).
It is not reasonable to think changing the boundary 
conditions will have much effect on the
free energy, a bulk quantity. Therefore, the transition 
must still be symmetry-breaking, even with open
boundary conditions. But what is the order parameter and 
what symmetry is broken in this case?  
The current study began as a hunt
for such an order parameter, as well as an order parameter 
for the attached first-order transition
in the SU(2) fundamental-adjoint plane. Whether or not the 
latter is a symmetry-breaking transition is a
subject of some controversy. 
Indirect evidence from the scaling of the latent heat with 
the size of the metastability region seems to 
indicate it is of the symmetry-breaking type\cite{myfa}.
Clear identification of a bulk symmetry-breaking order 
parameter associated with this transition would 
provide powerful confirmation of this idea. It is important 
because a symmetry-breaking first-order transition
must end in a tricritical point (as opposed to an ordinary 
critical point). The transition then must continue
from the tricritical point as a higher-order bulk transition 
which continues all the way through the phase plane,
separating the plane into distinct symmetry-broken and 
symmetry-unbroken regions. The strong-coupling
confining phase could not, in this case, be analytically 
connected to the weak coupling phase, but would
be everywhere separated by a phase transition, 
contrary to the conventional picture.

It is well known that, although physical observables are 
gauge invariant, many important features in 
understanding the behavior of gauge theories are only 
observable in a fixed gauge. It is also well known,
from Elitzur's theorem, that a local gauge symmetry cannot be 
spontaneously broken \cite {Elitzur}. The logic 
followed in this paper is to fix the gauge enough 
to invalidate Elitzur's theorem, and see if any portion of
the remaining exact partially-global 
gauge symmetry is broken spontaneously. If so, then one 
has found a broken symmetry and
associated order parameter. This is somewhat 
reminiscent of symmetry breaking in the standard model
except that there are no fields here besides the gauge fields.  
The gauge chosen is an axial gauge where
all links on a tree (connected set of links with no loops) 
are set to one. The tree chosen is
not the usual maximal tree, but rather the maximal tree 
with the last line of links pointing
in the fourth direction left off, leaving $N$ disjoint trees.
This leaves N independent exact SU(2) gauge symmetries 
(on an $N^4$ lattice), which operate ``globally'' 
on 3-d hyperplanes of fixed $x_4$ (coordinate in fourth direction). 
It is these symmetries which
appear to be spontaneously broken at weak coupling and 
undergo a phase transition to an 
unbroken state at strong coupling. The order parameters 
are simply the layered-magnetizations
formed from averaging the SU(2) links pointing in the 
fourth direction that lie in each hyperplane.
Each hyperplane is in a sense acting like a 3-d O(4) 
Heisenberg model, linked to other intersecting
and adjacent 
hyperplanes through the gauge couplings. Indeed, in two 
dimensions, the SU(2) gauge theory in the
axial gauge actually splits into N non-interacting 
1-d O(4) Heisenberg models\cite{Kogut}. 
Because the links
in the opposite direction are all set to one, the 
four-link gauge couplings turn into two-link spin
couplings. The same thing still happens for the 4-1 
plaquettes of the 4-d theory, but not for the 
4-2 or 4-3 plaquettes, also connected to the 4-links. 
Nevertheless, it might still make sense to think 
the 4-d theory as a set of 3-d Heisenberg models with
additional interactions. 
These interactions may make the critical
behavior different from that of the corresponding Heisenberg model.

\section{Results and Analysis}
Writing each SU(2) link 
as $a_0+ia_1 \tau _1 +ia_2 \tau _2 +ia_3\tau _3$ subject to the 
constraint $\sum_{k=0}^3 a_k ^2 =1$, they can be 
pictured as unit 4-d vectors in the
color parameter space. These vectors, for links 
pointing in the fourth spatial direction,
are averaged 
over each $x_4 = \rm{const.}$ hyperplane, to 
obtain $N$ magnetization color-vectors 
$\vec{m}_i$, $i=1 \cdots N$,
normalized per link.
In the unbroken phase, these magnetizations sum 
to small random values consistent with
Gaussian fluctuations around zero. In the broken 
phase they take on larger values, which drift
slowly with Monte Carlo
time, fluctuating around a nonzero magnitude and 
avoiding values near zero. In finite-lattice
simulations one always has the problem that due 
to tunneling between broken vacua (the slow
drift mentioned above) the order parameter itself 
will average to zero also in the broken phase.
To distinguish the phases one can either monitor 
the magnitude of the magnetization or introduce
a small explicit symmetry breaking through an 
external field. Both approaches are shown here.

The phase transition is, of course, most 
visible when it is first order, so the first 
case to be considered is in the fundamental-adjoint 
plane with adjoint
coupling $\beta _ A = 1.5$. 
This is below the triple-point but in a place where
there is a fairly strong first-order transition 
with a jump of about $0.27$ in the
average plaquette across the transition\cite{myfa}. 
For the four-dimensional order parameter one must take 
into account the geometrical factor 
(from solid angle) that biases the distribution toward
larger magnitudes. In the unbroken phase, 
the distribution of magnetization magnitudes, $m$, is
expected to be a factor of $m^3$ times a 
Gaussian, $\exp (-m^2/2 \sigma _{m} ^2)$. To more easily
see the Gaussian behavior, the probability distribution $P(m)$ is 
obtained by histogramming, and
the quantity $P(m)/m^3$ is plotted in Fig.~1, 
for the cases $\beta = 1.00$, on the 
strong coupling side, and $\beta = 1.09$, 
on the weak coupling side of the transition, which
occurs around $\beta = 1.04$. Here $\beta$ 
is the fundamental coupling parameter.
In the first case the data fit well to 
the expected Gaussian
peaked at zero.  The value of $m$ for each 
bin is not taken at the center,
but at a value that would produce a flat 
histogram in an $m^3$ distribution, regardless
of bin-size choice. This is 
\begin{equation}
m^{3}_{\rm{bin}}=\frac{1}{4}\frac{(m_{2}^{4}-m_{1}^{4})}{(m_2 - m_1 )} .
\end{equation}
where $m_2$ and $m_1$ are the bin edges. This 
detail affects only the first couple of bins in the 
histograms shown, and is necessary to get good 
Gaussian fits in the symmetric phase. 

On the weak-coupling side of the transition, however, 
the distribution shows a definite non-zero peak,
indicative of spontaneous symmetry breaking. This 
data was taken on a $12^4$ lattice with 5000
equilibration sweeps followed by 50,000 measurement 
sweeps. The 12 magnetizations on the different
lattice layers behave identically, as they must 
due to translational symmetry
with periodic boundary conditions. Their magnetization directions
are random and they appear uncorrelated with each 
other, although detailed 
correlation analyses have not yet been performed. 
Each is treated as an independent measurement 
in the histogram. Fig.~2 shows a run which was first 
equilibrated at 1.09, with the beta value 
then abruptly changed to 1.01. One expects a period of metastability 
followed by an eventual tunneling to the symmetric phase. 
By watching the tunnelings of various quantities occur, one can
get an idea of how closely they are correlated.
In Fig.~2a, the plaquette and a sample of four of the 
twelve link magnetizations are shown. The
others behave similarly.
The unbreaking of the symmetries appears to occur
just as the plaquette tunnels to its lower value in 
the unbroken 
phase. Fig.~2b shows the same for the Polyakov loop, 
which also appears to enter the
restored-symmetry phase at the same time. The Polyakov 
loop may in some sense be measuring the 
product of the N magnetizations, but this relationship needs 
further exploration.
Simulations at $\beta=1.09$ and $\beta=1.0$ were also 
performed with an open boundary condition (BC)
in the fourth direction.
The results for layers located more than two lattice 
spacings from the open boundary are close
to those given above. For instance the 
average magnetization at $\beta=1.09$ with
periodic BC's was $0.149(2)$, and with open BC's, 0.146(3).  
However the histograms are somewhat different.
The histogram for open boundary conditions at this coupling
(again ignoring layers near the boundary)
picks up a second peak close to zero. Due to the $m^{-3}$ 
multiplying factor, only 
a few occurrences of near-zero magnetization can result in a peak here.
Examination of the runs shows a few brief excursions near
zero for some of the layers, which might be interpreted 
as ``tunneling attempts". Probably what has happened is that the open BC has
shifted the critical point up a bit to the point where
tunneling is beginning to occur.  A simulation at a slightly weaker coupling,
$\beta=1.3$, 
$\beta _A = 1.5$, shows a very definite symmetry-broken histogram (Fig.~3),
similar to Fig.~1b. Open boundary condition runs with the Wilson 
action (see Fig. 4c,d below) also give very similar histograms to periodic 
boundary condition runs at the same coupling. 
Thus boundary conditions 
are apparently not fundamentally important in this transition. The symmetry is
still defined and still breaks at the transition, supporting the 
bulk nature of the transition. It should be noted that only the relevant
boundary in
the 4th direction, where the symetry breaking is being examined, was opened in
these runs. The other directions had periodic boundary conditions.

Simulations were also run on the Wilson axis. Here a 
higher-order transition is expected.
No bulk transition has previously been seen here at all -- only
the Polyakov loop transition which has been interpreted as
a finite temperature transition, one which disappears 
on an infinite 4-d lattice. However, 
since a bulk symmetry is now seen to be breaking 
elsewhere in the fundamental-adjoint plane it must also be breaking here. 
Fig.~4 shows histograms for $12^4$ Wilson-axis
runs at $\beta = 2.4$, 2.9, 3.2, and a $16^4$ run at $\beta = 3.8$. 
The former is seen 
to be unbroken and 
the latter all broken. Also
shown (Fig.~5) are runs taken with an external field term 
of the form $-a_0 h$ for each fourth-dimension-pointing
link added to the action
(this term is not multiplied by $\beta$).
Here the $a_0$-component of magnetization is plotted as well 
as the magnitude.
Only a small external field is needed to order the magnetization
in the broken phase, which clearly extrapolates to a non-zero
value in the limit of zero field. In contrast, in the 
unbroken phase ($\beta=2.2$)
the magnetization extrapolates nearly linearly to zero.
In determining what constitutes a ``strong'' or ``weak'' field,
the magnitude of $h$ should be compared to $\beta$ since
both links and plaquettes range from -1 to 1. The 
transition is consistent
with being coincident with the deconfinement transition, which 
on a $12\times \infty ^3$  lattice is projected to
be at a $\beta _c$ of 2.64 \cite{lattice2004}. However, 
the location of both transitions
on the finite $12^4$ lattice is quite fuzzy, so a 
precise statement of coincidence 
is not possible at this time. It should be noted, 
however, that the layer spin 
magnetization breaks the Z2 Polyakov loop symmetry as 
well as the aforementioned gauge
symmetry when it takes a non-zero expectation value. 
Without the $Z2$ symmetry to protect
it, the Polyakov loop would be expected to also take 
an expectation value in this phase.
Therefore it is extremely doubtful that the Polyakov 
loop transition could lie at a weaker
coupling than the layered SU(2) transition. It could, 
however, lie at a stronger coupling, since
restoration of the gauge symmetry does not necessarily imply 
restoration of the Z2 Polyakov loop symmetry.

Generally, the ``finite lattice transition point'' can be 
identified with a peak in the susceptibility,
defined for the magnetization magnitude as \begin{equation}
\chi = N^3 (<m^2 >-<m>^2 ) .
\end{equation}
However, for this system, Fig.~6 shows the susceptibility 
continues to rise  
at the weak coupling side of the transition (the $12^4$ 
lattice is broken beyond about $\beta =2.6$ 
and the $16^4$ beyond 2.8). Below $\beta=2.6$ all lattices
give the same result, whereas at higher $\beta$ a large 
lattice-size dependence is seen, consistent with 
a possible divergence in the limit
of infinite lattice size.
Divergence without a peak is a possible indication of the system
being critical at every $\beta$ above the transition 
(a massless phase), as would occur at a
BKT-like transition. 
Indeed, $\chi$ for the 2-d XY model also does not peak 
at the critical point, but 
continues to rise in the low-temperature phase \cite{XY}, and
appears to be divergent everywhere there.
Finite size scaling expected for such a transition 
is $\chi \propto N^{2-\eta }$, where $\eta$ is defined by 
the correlation 
function in the weak-coupling phase which is 
$\propto (x-x')^{-\eta }$ \cite{Barber}.  Finite size scaling 
analysis between $8^4$, $12^4$ and $16^4$ 
lattices does not appear to be consistent with a power law. 
The $8^4$ lattice appears to have saturated 
and is probably simply 
too small to provide any information on 
asymptotic behavior. Such 
saturation is not a surprise, since $<m>$ is 
bounded by unity and its 
fluctuations are similarly bounded. 
Above $\beta = 3.8$, the $12^4$
and $16^4$ lattices are marginally consistent 
with $\eta = 1$, but the 
errors are large. 
One obtains $\eta = 0.84 \pm 0.18$, $1.25 \pm 0.21$, and 
$1.24 \pm 0.23$ at 
$\beta$ = 4.2, 4.5, and 5.0 respectively. If these are 
combined, the result is $1.08 \pm 0.12$. 
Clearly higher statistics and larger 
lattices will be needed to measure the critical 
exponents with reasonable accuracy, and indeed to verify the
nature of the criticality. 
If the system were scaling as the O(4) 3-d Heisenberg model, 
the prediction would 
be $\chi \propto N^{1.975}$ \cite{O4}, which 
appears to be ruled out. 
A peak would also be expected in this case.

Finally, the Binder parameter\cite{Binder}, 
\begin{equation}
U = 1-<m^4>/(3<m^2 >^2 )
\end{equation} 
is considered.
For a finite-order transition, curves of the Binder 
parameter should cross at the 
infinite lattice transition point, at least in the 
limit of large lattices (next
to leading corrections to scaling 
may make the smaller lattices miss the intersection somewhat).
For the infinite order BKT-like transition, the 
Binder parameters for different size
lattices do not cross, but merge, in the weak-coupling phase, 
because the system is
critical throughout this phase 
(not just at the critical point). Fig.~7 shows the
Binder parameter for this system on the three lattices. 
For an O(4) order parameter,
the expected value for the Binder parameter in the random phase 
is $\frac{1}{2}$, and $\frac{2}{3}$ in
a fully-ordered phase\cite{O3paper}. At criticality, 
the value should be a critical
value somewhere in between these, which, except for
higher order corrections, should
be the same on all lattices. For a BKT transition, which is 
critical at every coupling in the weak-coupling phase, 
the critical $U$
could be a function of $\beta$. At strong coupling, 
$U$ for the larger lattices moves
down strongly, consistent with approaching the expected 
value of $\frac{1}{2}$ in the symmetric
phase. $U$ rises rapidly up to around $\beta = 3.5$, 
after which it rises more slowly, with the larger lattices
always below the smaller. 
There is no evidence for crossing or for large lattices 
approaching $\frac{2}{3}$,
even far into the symmetry-broken region, as would be
the case for an ordinary finite-order transition.
The behavior is consistent with a massless phase 
beginning at a BKT-like transition which
has an infinite-lattice transition point 
somewhere around $\beta = 3.5$ to $4.0$.
However, there is still way too much finite-lattice dependence to 
determine the critical values from these data -- 
larger lattices will be needed.

Another quantity which seems to favor the 
existence of a BKT-like transition is
the (pseudo) specific heat. The SU(2) theory has a 
large non-singular peak in the 
specific heat around
$\beta = 2.25$ which does not grow with 
lattice size\cite{Engels}. Interestingly
this is precisely the behavior of the X-Y model 
specific heat \cite{XY,Ch}, which has a
large nonsingular peak 
displaced a long way from the critical point.
The essential singularity at the critical point is very 
soft and not visible in numerical simulations.
It is located where the specific heat begins 
its rise on the weak-coupling side of the peak.
The X-Y model also has a very
large finite-lattice shift in the critical point, 
due to the infinite-order scaling\cite{XY}.
This could also explain the large $\beta$-shift 
seen in the deconfinement transition of the SU(2) 
gauge theory. 

Some particulars of the simulations are as follows. 
Most runs were for 50,000 sweeps with 5000 equilibration sweeps
for the $8^4$ and $12^4$ lattices, and 10,000 equilibration sweeps 
for the $16^4$ lattice. A Metropolis Monte Carlo algorithm
was employed with acceptance kept between 0.3 and 0.7. Measurements
were taken after every sweep. Equilibration times were set to 
twice the time it appeared to take for observables to equilibrate.
The last observable to equilibrate was generally the Polyakov loop 
in the first direction, where all but one link are set to unity
in this gauge. It takes some time for the information to percolate
into that one link, definitely more than for the Polyakov loop
to equilibrate in the non-gauge-fixed case, 
but not an unreasonably long time. 
For $\beta=4.2$ and above, 100,000 measurement sweeps were taken. 
Runs with
an external field present (Fig.~5) were 
shorter, only 5000 measurement sweeps, because the magnetization
itself (as opposed to its moments) is fairly easily measured.

\section{Conclusion}
In conclusion, a symmetry breaking phase 
transition is seen on 3-d layers of
4-d SU(2) lattice gauge theory in a 
partially-fixed axial gauge. The symmetry broken
is the remaining exact SU(2) global 
gauge symmetry on each layer, with the order
parameter being the link magnetization 
in the unfixed direction on each 
perpendicular hyperplane. The transition also 
remains for the case of open boundary 
conditions. The lack of dependence on boundary conditions, 
the bulk nature of the order parameter (existing locally on all
layers rather than globally in the fourth 
direction or on a single surface) 
and the apparent 
coincidence of the transition with the bulk
first-order transition in the fundamental-adjoint 
coupling plane, strongly
support this being a bulk transition. Finite-size 
scaling of the susceptibility
and Binder parameter suggest a BKT-like transition 
with the system being in a 
massless phase for all couplings smaller than the 
critical coupling, including
the continuum limit.

It is interesting to consider if this symmetry 
breaking has implications for the 
continuum theory, since the continuum limit is 
in the symmetry-broken phase.
Normally no part of the gauge symmetry is thought to be 
spontaneously broken in the continuum pure-gauge 
theory. However, this work would seem to indicate 
that the gauge fields themselves
can and do break a portion of the (partially global) 
gauge symmetry remaining after
partial gauge fixing. The gauge fields observed in 
nature could be the Goldstone
bosons associated with this symmetry breaking. It 
is not clear whether this has any
physical implications or whether it is simply a 
different way of looking at
the continuum gauge theory resulting from the 
choice of gauge.  

It is also very interesting to consider what 
would happen if the
gauge fixing were further relaxed, by fixing 
aong only the first two directions.
Such an investigation is being planned.
This would leave $N^2$ exact SU(2) symmetries 
which are global
on two-dimensional planes of fixed ($x_3$,$x_4$). 
There would 
be two order parameters, the link magnetizations 
in both the 3
and 4 directions. However, it is unclear whether 
these could 
break spontaneously or not, due to the 
two-dimensionality of the layers.
In a true two-dimensional system, the 
Mermin-Wagner theorem prevents
spontaneous breaking of a continuous symmetry. 
Here, however,
the two-dimensional layers are linked to others. 
It is possible that 
a phase transition on these underlying 
two-dimensional layers is
how BKT-like behavior finds its way into a 
four-dimensional theory.

\newpage
\begin{figure}[ht]
 \includegraphics[width=2.5in]{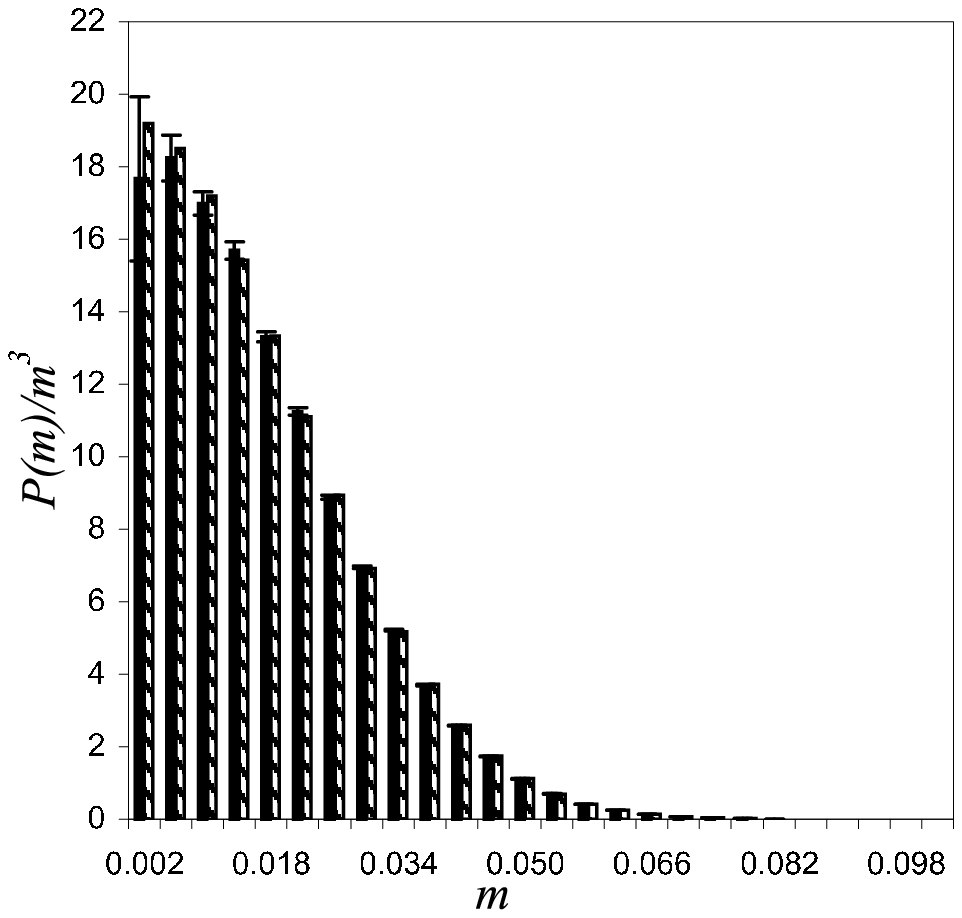} 
             \includegraphics[width=2.53in]{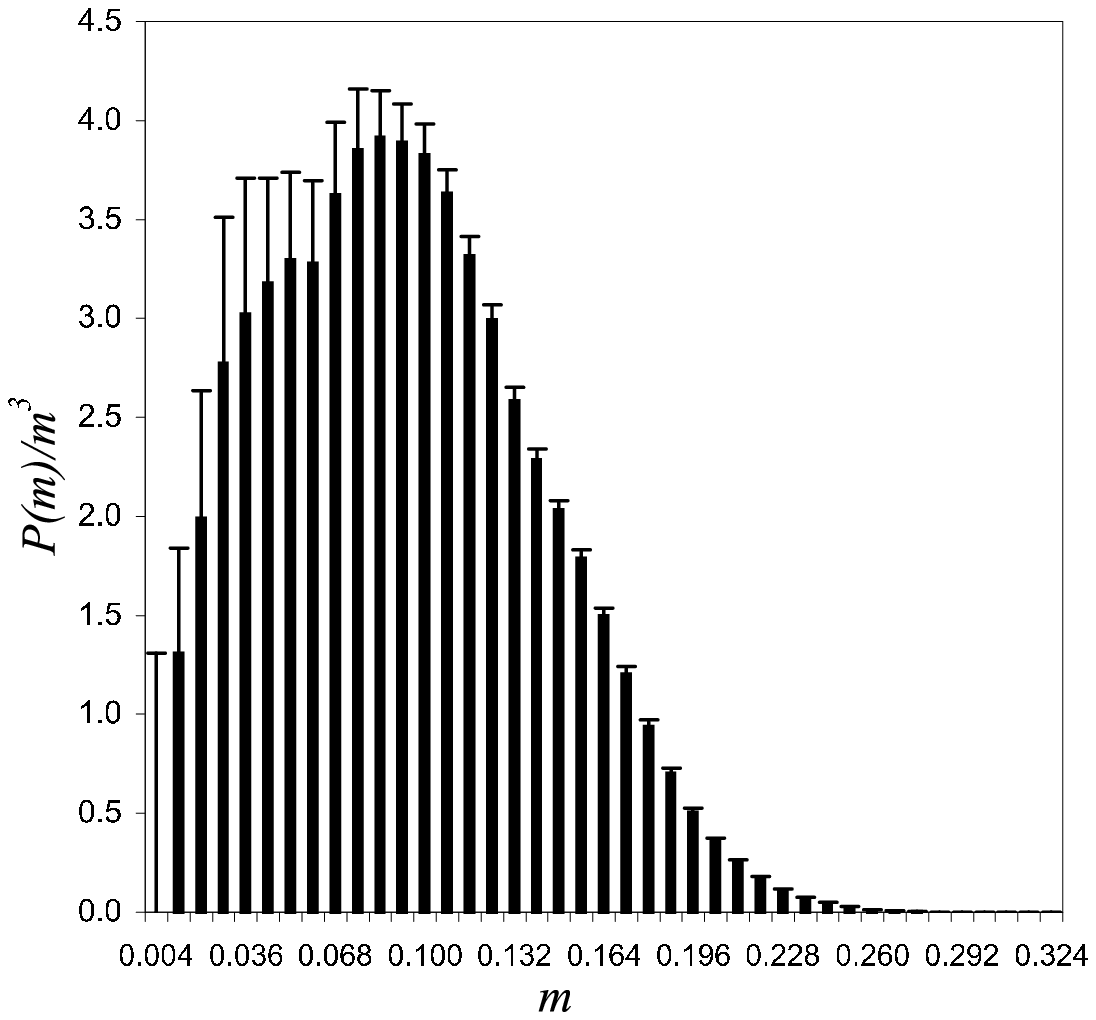} 
            \caption{Histograms for (a) $\beta _A = 1.5$, $\beta = 1.0$,
just on the strong coupling side and (b) $\beta _A =1.5$, $\beta= 1.09$, 
just on the weak coupling side of the first order transition in 
the fundamental-adjoint plane. Error bars are from binned fluctuations.
Striped bars in (a) are a fit to a Gaussian centered on zero. All histograms 
are arbitrarily 
normalized to integrate to unity. Labels on horizontal axis give middle of
bin for every fourth bin, beginning with the first.}
          \label{fig1}
       \end{figure}
\noindent
\begin{figure}[ht]
             \includegraphics[width=2.5in]{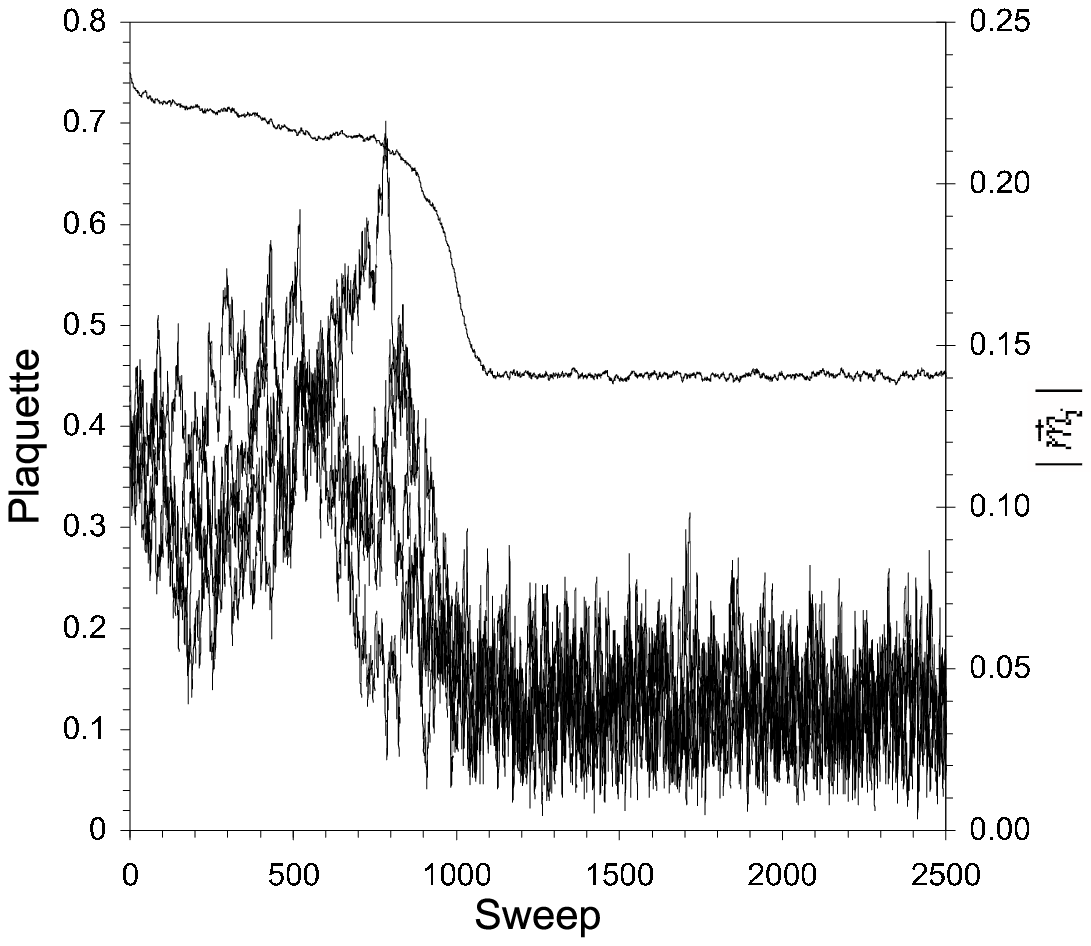} 
             \includegraphics[width=2.5in]{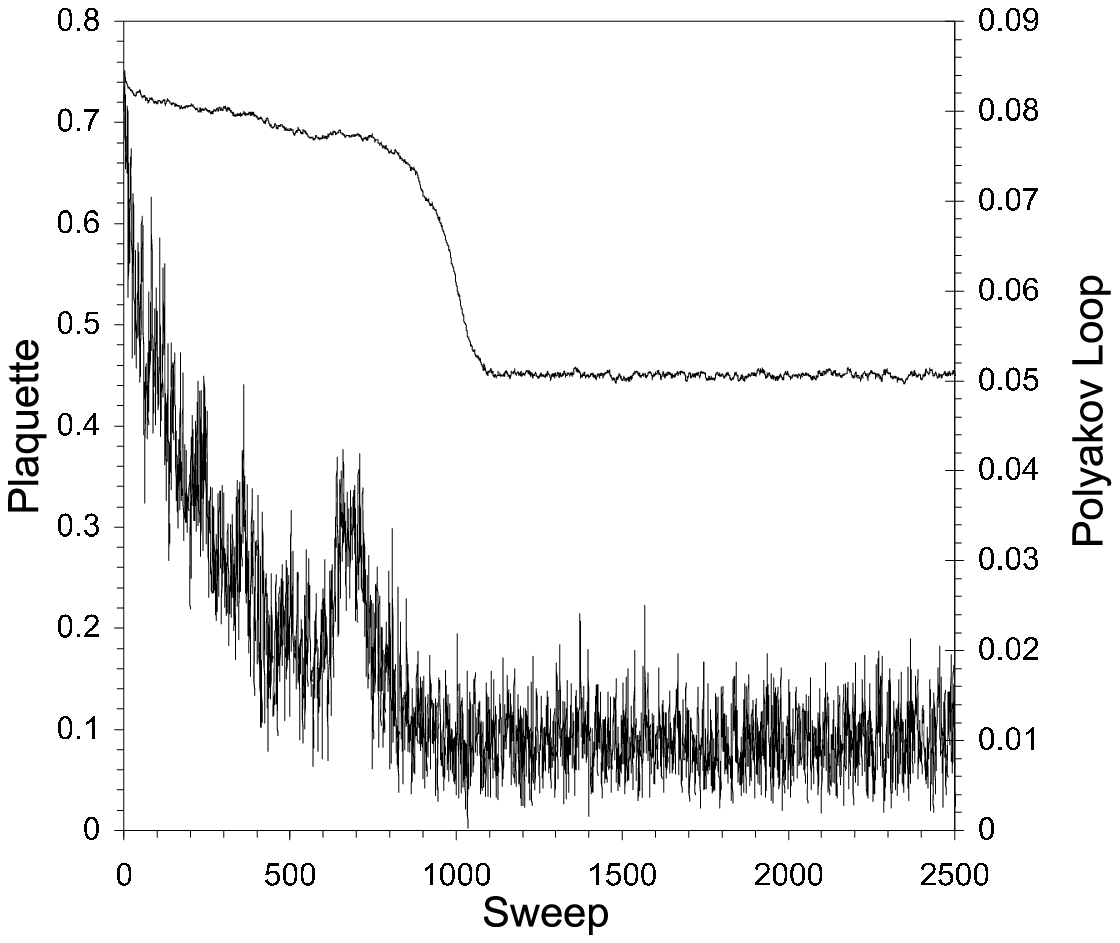} 
\caption{
Time history of a tunneling event on a $12^4$ lattice. A lattice
originally equilibrated at $\beta _A = 1.5$, $\beta=1.09$, was
subjected to a sudden change in fundamental coupling to $\beta = 1.01$
at time labeled zero in graphs. Upper curve is average plaquette in both 
graphs. Lower curves in (a) are average link magnetizations for a sample
of four of the twelve layers. Lower curve in (b) is the 
average Polyakov loop.}
\label{fig2}
\end{figure}
\noindent
\begin{figure}[ht]
             \includegraphics[width=4in]{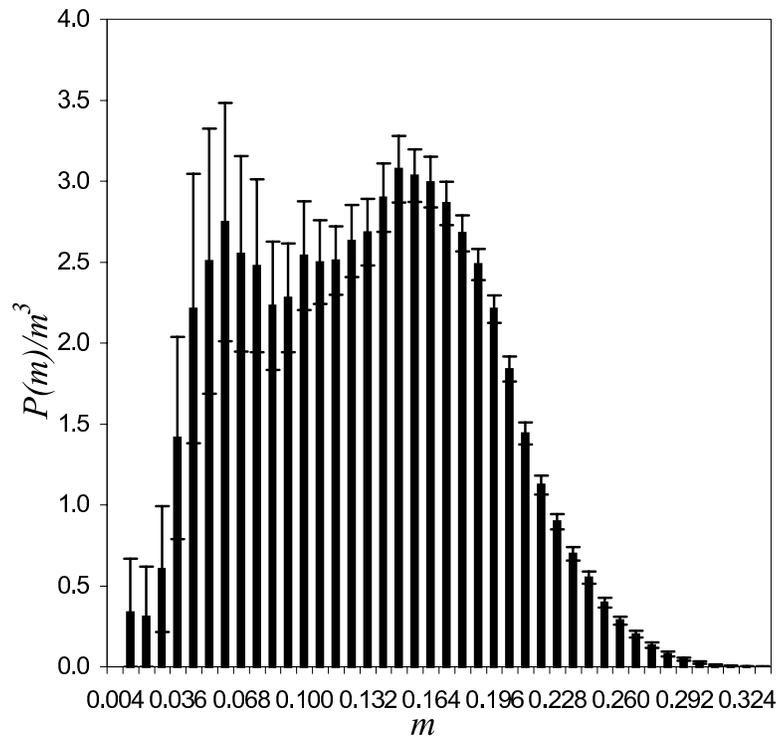} 
\caption{Histogram for $\beta _A = 1.5$, $\beta = 1.3$ with open boundary
conditions. The two closest layers on both sides of the boundary (a total of
four layers) are excluded. The first bin is empty.}
\label{fig3}
\end{figure}
\noindent
\begin{figure}[ht]
             \includegraphics[width=2.5in]{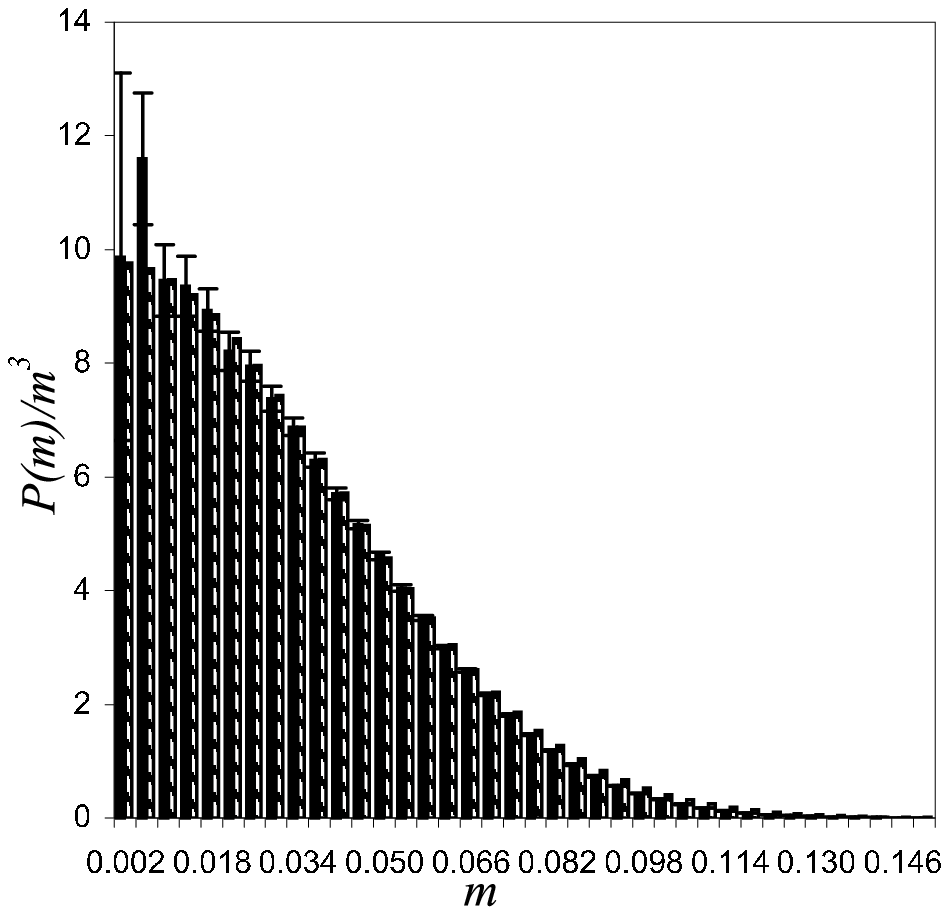} 
             \includegraphics[width=2.5in]{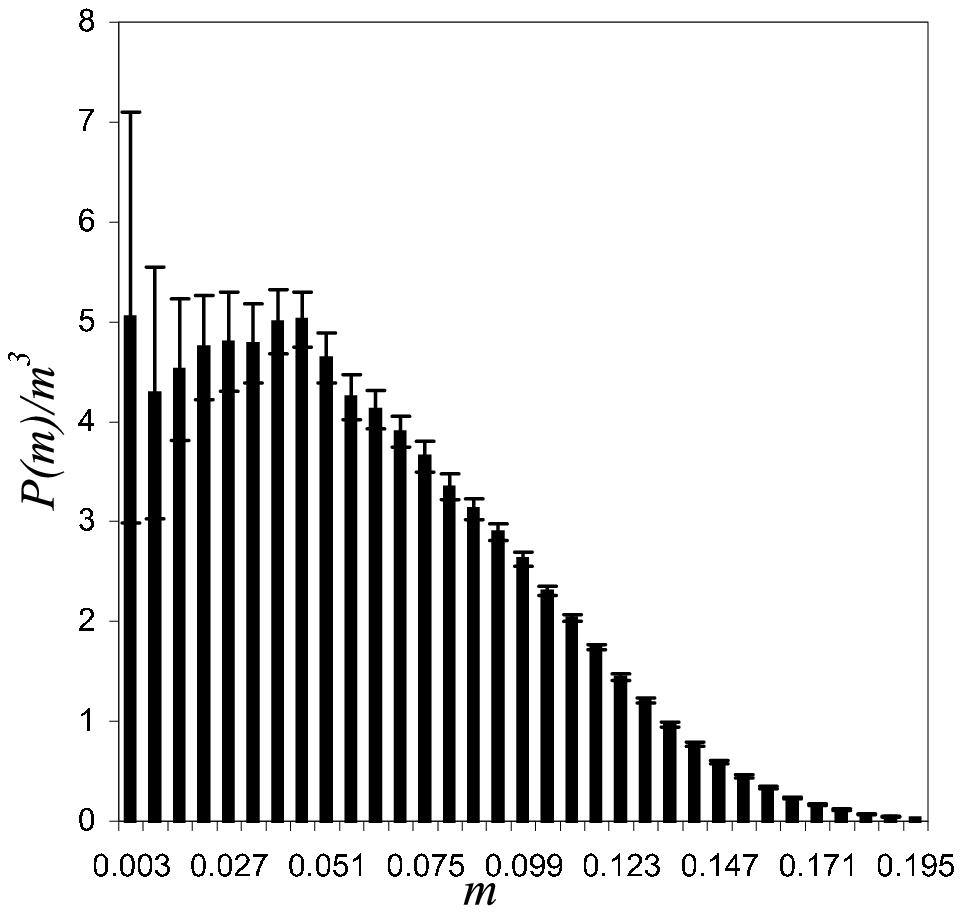} 
            \includegraphics[width=2.5in]{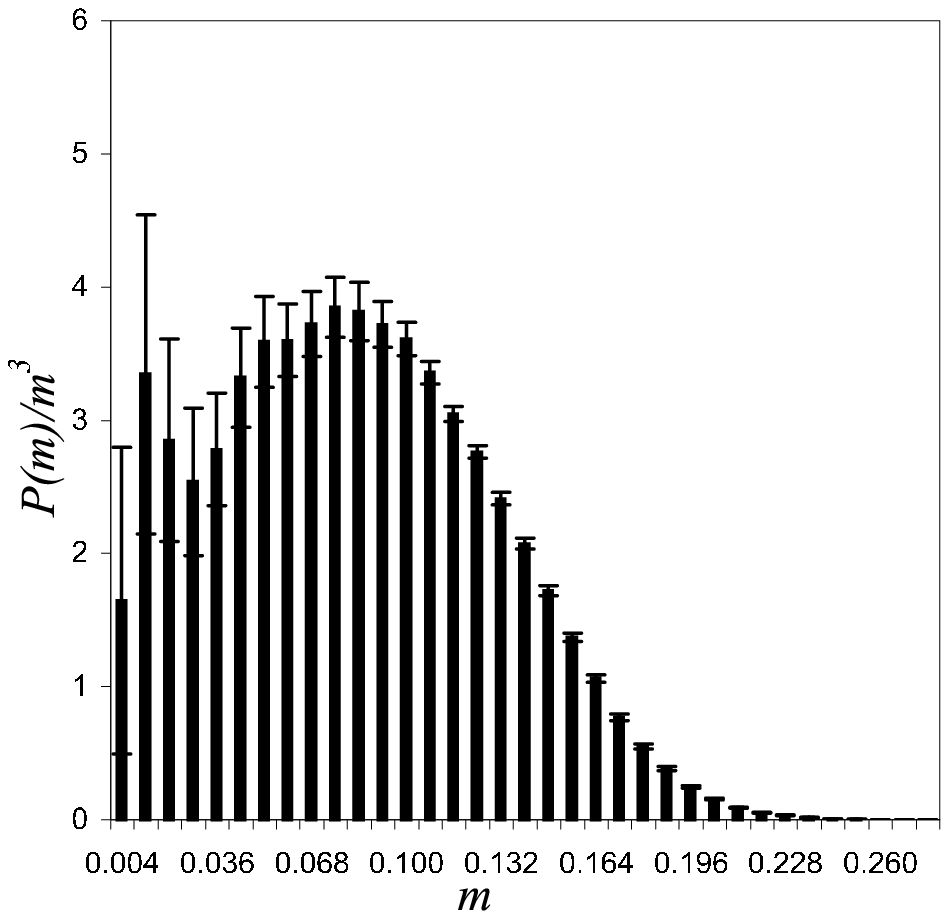} 
             \includegraphics[width=2.5in]{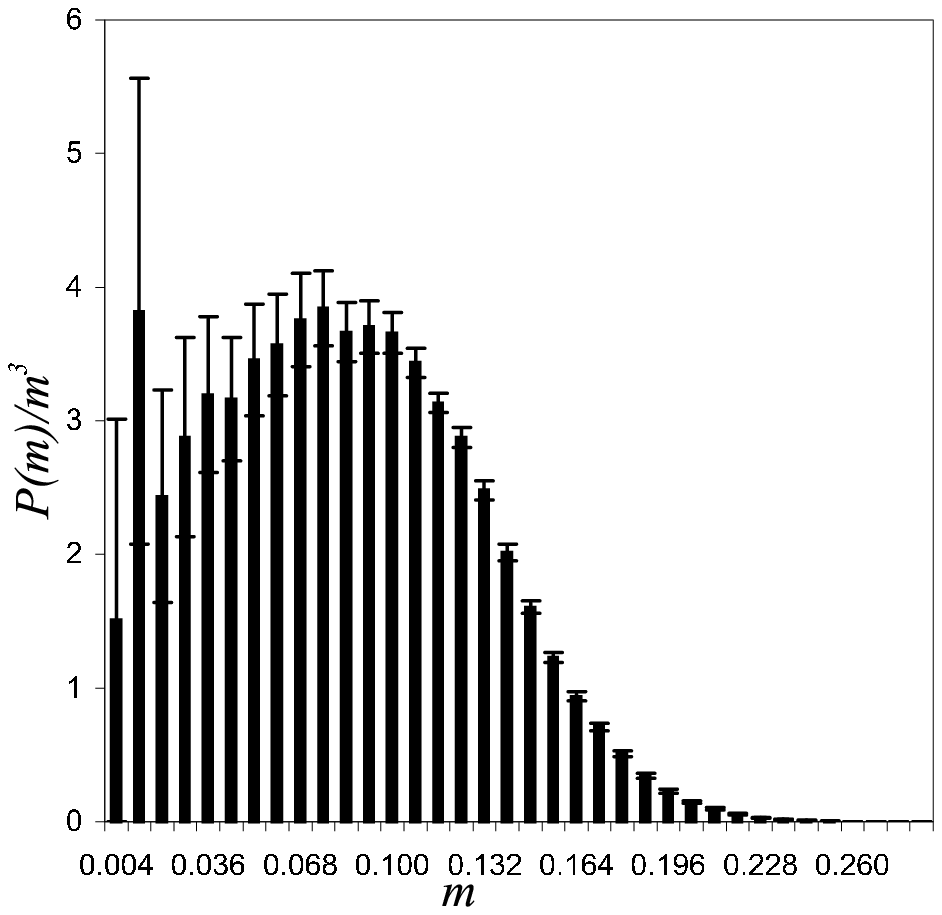} 
          \includegraphics[width=2.5in]{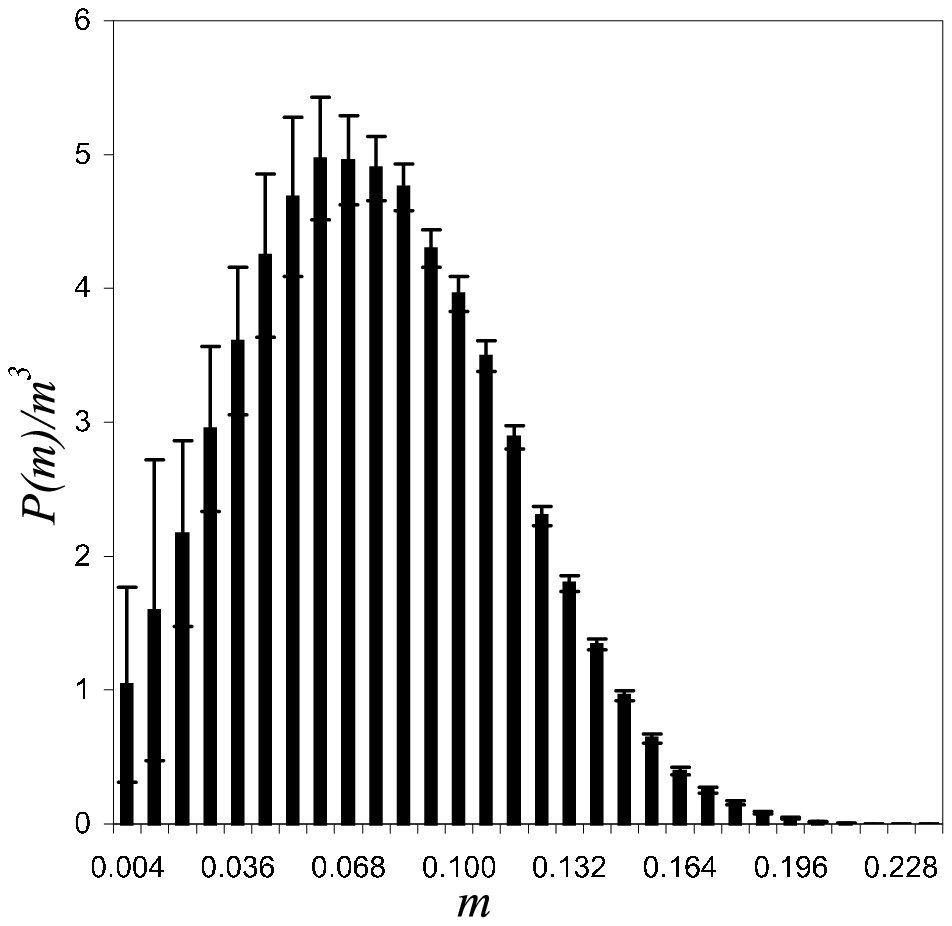} 
\caption{Histograms for Wilson-axis simulations at (a) $\beta = 2.4$,
(b) $\beta = 2.9$, (c) $\beta = 3.2$, (d) $\beta = 3.2$ with open 
boundary conditions, and (e) $\beta = 3.8$ on a $16^4$ lattice.
Other runs are on a $12^4$ lattice and all but (d) are with periodic
boundary conditions. Striped bins in (a) are a fit to a Gaussian
centered on zero. Fit is to the first 12 bins. Extreme tail of data in (a)
is somewhat lower than Gaussian would predict.}
\label{fig4}
\end{figure}
\noindent
\begin{figure}[ht]
             \includegraphics[width=4in]{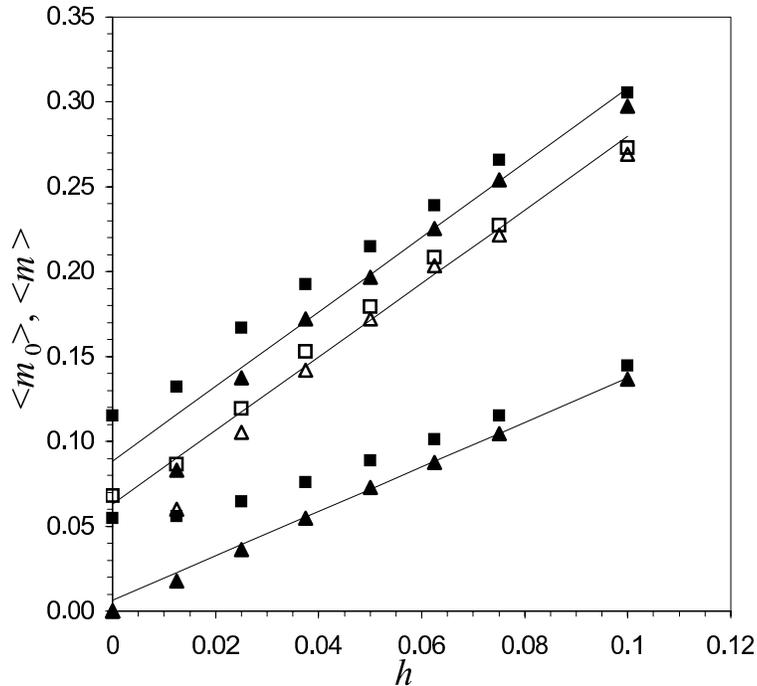} 
\caption{Effect of imposing an external field, $h$,
coupled to 0-component
of magnetization, here called $m_0$ and plotted as triangles. 
Also shown is the magnetization magnitude, $m$, plotted as squares.
Solid symbols are for $12^4$ lattice and open are for $16^4$. Lower 
data set is for $\beta=2.2$, and upper two are for $\beta=2.9$ on
the two different lattices.  Lines are linear fits to $<m_0 >$ in
the range $0.0375 \leq h \leq 0.075$ where this quantity is nearly 
linear.  The $\beta = 2.9$ data clearly extrapolate to  non-zero values
at zero external field,
which, for the $16^4$ lattice, is close to the zero-field value of $<m>$.
This is the expected hysteresis curve for a broken symmetry.
The 0-component of magnetization for $\beta=2.2$ extrapolates
to a point consistent with zero, as expected in the symmetric phase.}
\label{fig5}
\end{figure}\newpage
\noindent
\begin{figure}[ht]
             \includegraphics[width=4in]{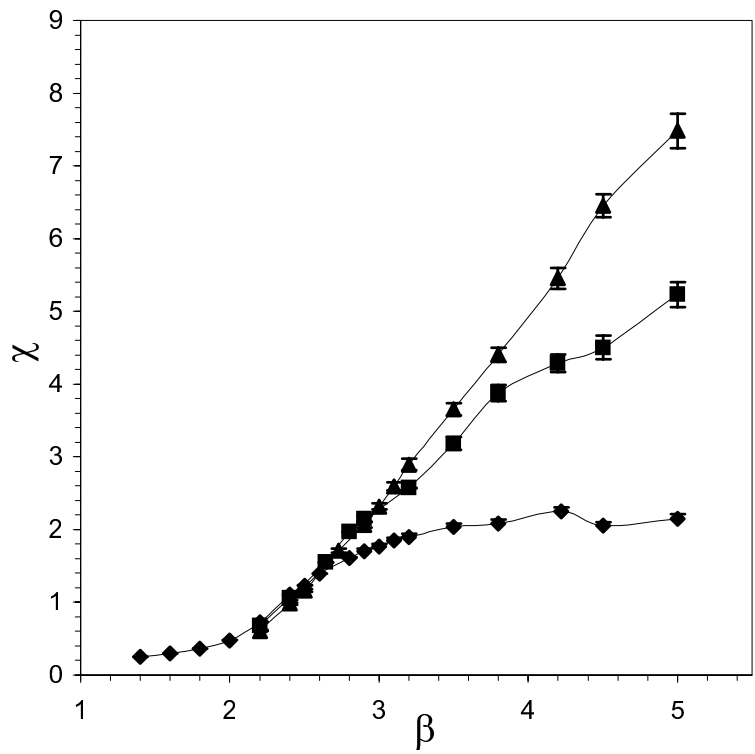} 
\caption{Susceptibility vs. $\beta$ for $8^4$ (diamonds),
$12^4$ (squares) and $16^4$ (triangles) lattices.}
\label{fig6}
\end{figure}\newpage
\noindent
\begin{figure}[ht]
             \includegraphics[width=4in]{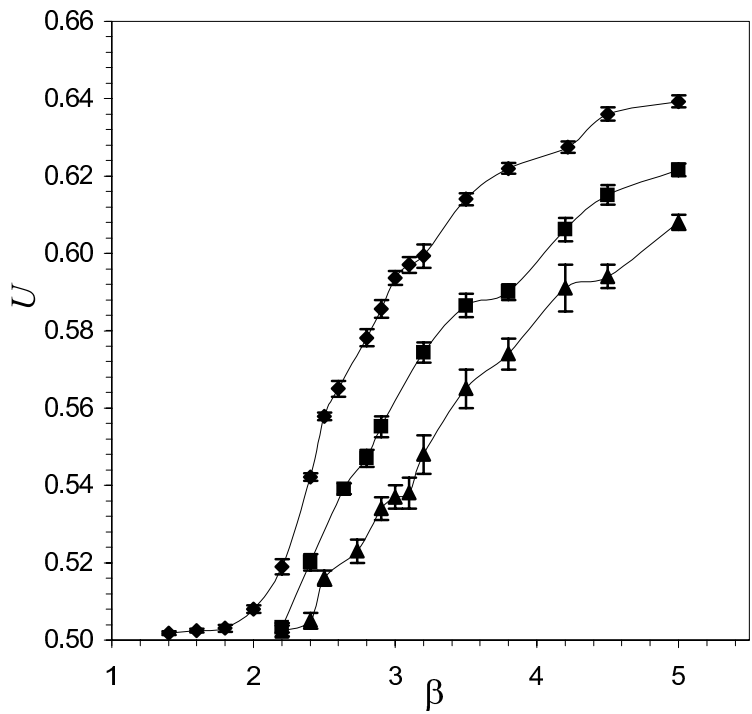}   
\caption{Binder parameter for $8^4$ (diamonds), $12^4$ (squares),
and $16^4$ (triangles) lattices. Prediction for large lattices in the
symmetric phase is 0.5 and in a broken phase for a finite-order transition
is 0.667. Critical values are expected to lie between these 
two extremes.}
\label{fig7}
\end{figure}
\end{document}